\documentclass{aa}

\usepackage{epsf}
\usepackage{graphics}

\begin{document}

\def\clipfig#1{\def\lbracket{[}\def\testit{#1}%
    \ifx\testit\lbracket\let\next=\optclipfig\else\let\next=\stdclipfig\fi%
    \next{#1}}
%
\newcommand {\hclipfig} [7] {\clipfig[#7]{#1}{#2}{#3}{#4}{#5}{#6}}
%
\def\usemodepsfig {\global\def\cfmode{x}\typeout{*** set clipfig to PSFIG mode ***}}
\def\usemodeepsf  {\global\def\cfmode{}\typeout{*** set clipfig to EPSF mode ***}}
\def\useunitmm    {\global\def\cfunit{x}\typeout{*** set clipfig to use mm as unit ***}}
\def\useunitcm    {\global\def\cfunit{}\typeout{*** set clipfig to use cm as unit ***}}
\def\clipfigsettings {\ifx\cfmode\empty\def\ccfmode{EPSF }\else\def\ccfmode{PSFIG }\fi%
    \ifx\cfunit\empty\def\ccfunit{cm }\else\def\ccfunit{mm }\fi%
    \typeout{*** current clipfig settings: \ccfmode mode, using \ccfunit as unit ***}}
%
%
%
%
\def\stdclipfig#1#2#3#4#5#6{\ifx\cfmode\empty%
    \let\next=\eclipfig\else\let\next=\pclipfig\fi%
    \next{#1}{#2}{#3}{#4}{#5}{#6}}
\def\optclipfig#1#2]#3#4#5#6#7#8{\ifx\cfmode\empty%
    \let\next=\ehclipfig\else\let\next=\phclipfig\fi%
    \next{#3}{#4}{#5}{#6}{#7}{#8}{#2}}
%
%
%
\newcommand {\pclipfig}[6] {\ifx\cfunit\empty%
        \psfig{figure=#1.ps,width=#2cm,bbllx=#3cm,bblly=#4cm,bburx=#5cm,%
           bbury=#6cm,clip=}\else%
        \psfig{figure=#1.ps,width=#2mm,bbllx=#3mm,bblly=#4mm,bburx=#5mm,%
           bbury=#6mm,clip=}\fi}
\newcommand {\phclipfig}[7] {\ifx\cfunit\empty%
        \hspace{#7cm}\psfig{figure=#1.ps,width=#2cm,bbllx=#3cm,bblly=#4cm,%
           bburx=#5cm,bbury=#6cm,clip=}\else%
        \hspace{#7mm}\psfig{figure=#1.ps,width=#2mm,bbllx=#3mm,bblly=#4mm,%
           bburx=#5mm,bbury=#6mm,clip=}\fi}
%
%
%
\newcommand {\eclipfig}[6]{%
  \ifx\cfunit\empty\epsfxsize=#2cm\else\epsfxsize=#2mm\fi%
  \epsfclipon\epsfverbosetrue%
  \cfcmtopspts{#3}\cfllxi=\cftempi\cfllxf=\cftempf%
  \cfcmtopspts{#4}\cfllyi=\cftempi\cfllyf=\cftempf%
  \cfcmtopspts{#5}\cfurxi=\cftempi\cfurxf=\cftempf%
  \cfcmtopspts{#6}\cfuryi=\cftempi\cfuryf=\cftempf%
  \def\cfstra{\number\cfllxi.\number\cfllxf}%
  \def\cfstrb{\number\cfllyi.\number\cfllyf}%
  \def\cfstrc{\number\cfurxi.\number\cfurxf}%
  \def\cfstrd{\number\cfuryi.\number\cfuryf}%
  \hbox{\epsfbox[{\cfstra} {\cfstrb} {\cfstrc} {\cfstrd}]{#1.ps}}}
\newcommand {\ehclipfig}[7]{%
  \ifx\cfunit\empty\epsfxsize=#2cm\else\epsfxsize=#2mm\fi%
  \epsfclipon\epsfverbosetrue%
  \cfcmtopspts{#3}\cfllxi=\cftempi\cfllxf=\cftempf%
  \cfcmtopspts{#4}\cfllyi=\cftempi\cfllyf=\cftempf%
  \cfcmtopspts{#5}\cfurxi=\cftempi\cfurxf=\cftempf%
  \cfcmtopspts{#6}\cfuryi=\cftempi\cfuryf=\cftempf%
  \def\cfstra{\number\cfllxi.\number\cfllxf}%
  \def\cfstrb{\number\cfllyi.\number\cfllyf}%
  \def\cfstrc{\number\cfurxi.\number\cfurxf}%
  \def\cfstrd{\number\cfuryi.\number\cfuryf}%
  \ifx\cfunit\empty\hspace{#7cm}\else\hspace{#7mm}\fi%
  \hbox{\epsfbox[{\cfstra} {\cfstrb} {\cfstrc} {\cfstrd}]{#1.ps}}%
  \vspace{-1mm}}
%
%
%
\newdimen\cfllxi \newdimen\cfllyi  \newdimen\cfurxi  \newdimen\cfuryi
\newdimen\cfllxf \newdimen\cfllyf  \newdimen\cfurxf  \newdimen\cfuryf
\newdimen\cftemp \newdimen\cftempi \newdimen\cftempf
\newdimen\cfpspoint \cfpspoint=1bp
%
%
%
\newcommand{\cfcmtopspts}[1]{\ifx\cfunit\empty%
  \cftemp=#1cm\else\cftemp=#1mm\fi%
  \multiply\cftemp10\divide\cftemp\cfpspoint%
  \cftempf=\cftemp\divide\cftemp10\cftempi=\cftemp\multiply\cftemp10%
  \advance\cftempf-\cftemp}
%
%
\def\cfmode{}\def\cfunit{}\clipfigsettings
%

\useunitmm


\newcommand{\lb}{$\lambda$}
\newcommand{\sm}[1]{\footnotesize {#1}}
\newcommand{\inft}{$\infty$}
\newcommand{\vlv}{$\nu L_{\rm V}$}
\newcommand{\lv}{$L_{\rm V}$}
\newcommand{\lx}{$L_{\rm x}$}
\newcommand{\lsoft}{$L_{\rm 250eV}$}
\newcommand{\lhard}{$L_{\rm 1keV}$}
\newcommand{\vlsoft}{$\nu L_{\rm 250eV}$}
\newcommand{\vlhard}{$\nu L_{\rm 1keV}$}
\newcommand{\vlir}{$\nu L_{60\mu}$}
\newcommand{\ax}{$\alpha_{\rm x}$}
\newcommand{\aopt}{$\alpha_{\rm opt}$}
\newcommand{\aoxs}{$\alpha_{\rm oxs}$}
\newcommand{\aoxh}{$\alpha_{\rm oxh}$}
\newcommand{\airhard}{$\alpha_{\rm 60\mu-hard}$}
\newcommand{\aoxsoft}{$\alpha_{\rm ox-soft}$}
\newcommand{\aio}{$\alpha_{\rm io}$}
\newcommand{\aixs}{$\alpha_{\rm ixs}$}
\newcommand{\aixh}{$\alpha_{\rm ixh}$}
\newcommand{\hb}{H$\beta_{\rm b}$}
\newcommand{\nh}{$N_{\rm H}$}
\newcommand{\nhgal}{$N_{\rm H,gal}$}
\newcommand{\nhfit}{$N_{\rm H,fit}$}
\newcommand{\ale}{$\alpha_{\rm E}$}
\newcommand{\cts}{$\rm {cts\,s}^{-1}$}
\newcommand{\pl}{$\pm$}
\newcommand{\kev}{\rm keV}
\newcommand{\rb}[1]{\raisebox{1.5ex}[-1.5ex]{#1}}
\newcommand{\ten}[2]{#1\cdot 10^{#2}}
\newcommand{\msun}{$M_{\odot}$}
\newcommand{\dM}{\dot M}
\newcommand{\dMM}{$\dot{M}/M$}
\newcommand{\dMedd}{\dot M_{\rm Edd}}
\newcommand{\kms}{km\,$\rm s^{-1}$}

\thesaurus{03(02.01.2; 11.01.2; 11.14.1; 11.19.1; 11.17.4)}

\title{New bright soft X-ray selected ROSAT AGN}
\subtitle{II. Optical emission line properties
\thanks{Based in part on
observations at the European Southern Observatory La Silla (Chile) with
the 2.2m telescope of the Max-Planck-Society during MPG and ESO time.
All spectra can be retrieved via CDS anonymous FTP 130.79.128.5.
}
}
\author{D. Grupe\inst{1, 2, }\thanks{
Guest Observer, McDonald Observatory,
University of Texas at Austin},
K. Beuermann\inst{1,2},
K. Mannheim\inst{2}
\and H.-C. Thomas\inst{3}
}
\offprints{\\ D. Grupe (dgrupe@xray.mpe.mpg.de)}
\institute{
MPI f\"ur extraterrestrische Physik, Giessenbachstr., 85748 Garching, Germany
\and Universit\"ats-Sternwarte, Geismarlandstr. 11, 37083 G\"ottingen,
Germany
\and MPI f\"ur Astrophysik, Karl-Schwarzschildstr. 1, 85748 Garching, Germany
}
\date{received 22.May 1998; accepted 3 August 1999}
\maketitle
\markboth{D. Grupe et al.: Optical Emission Line properties of ROSAT AGN}{ }

\begin{abstract}
We present the emission line properties of a sample of 76 bright soft
X-ray selected ROSAT Active Galactic Nuclei. All optical counterparts
are Seyfert 1 galaxies with rather narrow permitted lines, strong
optical FeII line blends, and weak forbidden lines. By selection, they
also have steep soft X-ray spectra when compared with typical Seyfert
1 galaxies. We discuss possible origins of these peculiar trends
employing detailed correlation analyses, including a Principal
Component Analysis. The optical spectra are presented in the Appendix.
\end{abstract}
\section{Introduction}

The optical/UV/soft X-ray bump has turned out to be a common property
of most Seyfert 1 type Active Galactic Nuclei (AGN).  With ROSAT
(Tr\"umper 1983) numerous AGN have been found that show a soft X-ray
excess, commonly believed to be the high energy part of the ``Big Blue
UV/Soft X-ray Bump''.

Before ROSAT, a study of soft X-ray selected AGN had to rely on
serendipitous observations, notably with the EINSTEIN Image
Proportional Counter (IPC). In a sample of 53 AGN from the Ultra Soft
Survey (USS) of C\'ordova et al. (1992), Puchnarewicz et al. (1992)
found preferentially AGN with relatively narrow permitted emission
lines and strong FeII emission. About half of their sample had
FWHM(H$\beta)~<~ 2000$~\kms, confirming the earlier finding that
X-ray selected AGN tend to show relatively narrow permitted lines
(Stephens 1989). Such AGN were defined by Osterbrock \& Pogge (1985)
as ``narrow-line'' Seyfert 1 galaxies (NLSy1).  
%
%
Boller et al. (1996) found vice versa that optically selected NLSy1
galaxies show very steep X-ray spectra, confirming the physical
relation between these properties (see also Boroson \& Green 1992,
Brotherton 1996, Laor et al. 1994, 1997, Brandt \& Boller 1998).

The present sample of 76 AGN results from the optical identification
of the brightest previously unknown soft X-ray sources at high
galactic latitude (Thomas et al. 1998, Beuermann et al. 1999) found in
the ROSAT All-Sky Survey, RASS (Voges 1993, Voges et al. 1999). By
definition, it excludes well-known sources\footnote{The initial
selection criteria of the sample were a total PSPC count rate $> 0.5$
cts/s, a negative hardness ratio HR1 of the spectral energy
distribution, a source location at galactic latitudes
$|b|~>20^{\circ}$, and sufficient previous knowledge of the optical
counterpart (see the identification papers by Thomas et al. 1998 and
Beuermann et al. 1999). Based on the final (and different) count rates
given the RASS Bright Source Catalog (Voges et al. 1999), the sample
is no longer complete in a flux-limited sense.}.

In Paper I (Grupe et al. 1998a)
we described the continuum properties of the 76 AGN
contained in the sample.  The mean soft X-ray spectral slope is
steeper ($<\alpha_{\rm X}>$ = 2.1, $F_{\nu} \propto \nu^{-\alpha}$)
than found in other AGN studies. At optical wavelengths, the soft AGN
have significantly bluer spectra than a comparison sample comprised of
AGN with a canonical, much harder X-ray spectrum, whereas the slope
between 5500~\AA~ and 1~keV is the same.  This implies stronger Big
Blue Bump emission relative to an underlying continuum.  The blueness
of the optical spectra increases with the softness of the X-ray
spectra {\it and} with the luminosity, saturating at an approximate
$F_\nu\propto \nu^{+0.3}$ spectrum and consistent with most of the Big
Blue Bump emission originating in an (comptonized) accretion disk and
stretching from optical wavelengths to the soft X-ray regime.

In this paper we investigate the properties of the optical emission
lines of the bright soft X-ray selected AGN and discuss their
connection with X-ray properties.  Results are compared with a control
sample of hard X-ray selected ROSAT AGN (see Grupe 1996 and Paper I) 
with optical data
taken from the literature and ROSAT Position Sensitive Proportional Counter 
(PSPC, Pfeffermann et al. 1986)
pointed observations
retrieved from the ROSAT public archive at the ROSAT Science Data
Center at MPE Garching.  Sect.~\ref{obs} deals with the observations
and data reduction.  Several diagnostically important correlations
between emission-line and continuum properties are noted in
Sect.~\ref{corr}.  Finally, we discuss the results in
Sect.~\ref{dis} addressing the issue of the origin of the difference
between Seyfert 1 and NLSy1 galaxies. In the Appendix we present the
optical medium-resolution spectra of the AGN as well as the FeII
subtracted spectra.

Throughout the paper, luminosities are calculated assuming a Hubble
constant of \mbox{$H_0 = 75$\,km\,s$^{-1}$Mpc$^{-1}$} and a deceleration
parameter of $q_0 = 0$.

\section{\label{obs} Observations and data reduction}
\subsection{Observations}

The statistical analysis of the RASS X-ray data was presented in Paper
I. There, we also discussed the overall spectral energy distribution
based on the X-ray data and the optical continuum inferred from the
spectrophotometry presented here. The optical continuum slope is
defined between rest-frame wavelengths of 4400 and 7000 \AA.

We have performed medium resolution ($\sim 5$\AA~ FWHM) optical
spectroscopy for all AGN of the sample, using the 2.2m MPI/ESO
telescope at La Silla, Chile and the 2.1m telescope at McDonald
Observatory of the University of Texas at Austin.
The flux-calibrated spectra of all 76 ROSAT AGN summarized in
\ref{opt-spec} and spectral parameters listed in Table \ref{opt-list}.
The positions quoted for the optical counterparts were derived from
the Digitized Sky Survey. Table \ref{opt-list} also provides
information on the instrumentation used, the exposure times, etc.
The spectra are published electronically at the anonymous FTP 130.79.128.5.

\subsection{\label{feii-sub} FeII subtraction}

All Seyfert 1 galaxies in the soft X-ray AGN sample show optical FeII
emission. The optical FeII emission is dominated by two blends,
between 4435 - 4700 \AA~ in the blue, and between 5070 - 5600 \AA~ in
the red.  Weaker highly ionized iron lines like [FeVII]$\lambda$5159
or $\lambda$5721 or even the HeII$\lambda$4686 line are blended with
this emission. These blends also contaminate strong lines such as
[OIII]$\lambda\lambda$4959,5007. In order to reliably measure line
fluxes and to determine the strength of the FeII emission it is
necessary to appropriately model the FeII complexes. We adopt the
method described by Boroson \& Green (1992) and used the FeII template
of I\,Zw\,1 given in their paper for the wavelength range $\sim$
4400-6000\AA.  In order to correct also for the bluer part of the
spectrum, this template was extended towards shorter wavelengths by
using the relative FeII line intensities given by Phillips
(1978a,b). The whole template was wavelength-shifted according to the
redshift of the object spectrum and the individual FeII lines were
broadened to the FWHM of the broad H$\beta$ line by using a Gaussian
filter.  The template was scaled to match the line intensities of the
object spectrum and then subtracted. It was found most reliable to
scale the template by fitting the intensities of the FeII lines at
4924 and 5018 \AA.

In order to determine the rest-frame equivalent width of FeII, we
measured the flux in the scaled template between the rest wavelength
4250\AA~and 5880\AA, and the continuum flux density at 5050 \AA~in the
FeII subtracted spectra. All spectra can be accessed by anonymous FTP 
130.79.128.5.

\subsection{Other line measurements}

The FeII-subtracted spectra were used to measure the non-FeII line
properties. In these spectra, H$\beta$ is still a mixture of broad and
narrow components which we attempt to separate for further
analysis. In the spectra of some objects of subtype Sy1.5 this is
straightforward, in others both components are not easily
separable. In order to isolate the broad H$\beta$ component as best as
possible in all spectra, we proceed as follows. We use a template
built from the [OIII]$\lambda5007$ line which includes the asymmetry
often found in this and other narrow lines. The H$\beta$ line was then
synthesized from this narrow and a broader Gaussian component. The
latter is referred to as \hb. The FWHM(\hb) is what we consider as
representative of the Broad Line Region (BLR). We chose to define the
[OIII]$\lambda5007$/\hb~ ratio such that it refers to the broad
component \hb. Finally, the instrumental resolution was determined
from the FWHM of the night-sky lines and all quoted line widths were
corrected for the instrumental resolution.

Deriving the Balmer decrement H$\alpha$/H$\beta$ requires correction
of H$\alpha$ for the contributions of both, the narrow H$\alpha$
component and the [NII]$\lambda\lambda$6548,6584 lines. In many cases,
however, it proved difficult to isolate the broad H$\alpha$ component
this way. In order to correct for the contributions of
[NII]$\lambda\lambda$6548,6584 we assumed, therefore, that these lines
contribute 35\% of the flux in [OIII]$\lambda5007$, following Ferland
\& Osterbrock (1986). We refrained from separating broad and narrow
Balmer components and measured the Balmer decrement from the total
(broad and narrow) fluxes in H$\alpha$ and H$\beta$. These values of
the Balmer decrements do not differ appreciably from those of the
broad components because the relative contributions of the narrow
components to the Balmer line fluxes is usually small. Our Balmer
decrements are not necessarily representative of the narrow components.

\subsection{Continuum measurements}

Most of the V-magnitudes in Table \ref{opt-res} were derived from our
spectrophotometry. Although the spectral fluxes were corrected as
carefully as possible for slit losses in flux, the absolute fluxes are
uncertain by some 0.2 mag. For 22 southern objects, we determined more
accurate V-magnitudes from CCD photometry (see Thomas et al. 1998). In
this paper, we do not distinguish between the different sources of V,
because the intrinsic scatter in the data is larger than that produced
by errors in measuring the optical fluxes. The optical continuum slope
\aopt~ is measured between 4400-7000 \AA~ in the rest frame.  The
X-ray spectral slope \ax~ is defined between 0.2 and 2.0 keV (see
Paper I for more details).

\section{\label{corr} Results}

{\footnotesize
\begin{table*}
\begin{flushleft}
\caption{\label{opt-res} Optical results of the AGN. The
FWHM are given in \kms, equivalent width EW of FeII in \AA~in the
rest-frame, and the luminosity of \hb~ in Watt  }
\begin{tabular}{rlccr@{}l@{}rr@{}l@{}rccccccl}
\hline\noalign{\smallskip}
& & & & \multicolumn{6}{c}{FWHM} & EW & & log & log & log & log \\
\rb{No.} & \rb{Object} & \rb{V} & \rb{z} & 
\multicolumn{3}{c}{H$\beta$} & \multicolumn{3}{c}{[OIII]} & FeII &
\rb{$\rm \frac{H\alpha}{H\beta}$} & $\rm \frac{[OIII]}{H\beta}$ & 
$\rm \frac{FeII}{H\beta}$ & $\rm \frac{FeII}{[OIII]}$ 
& $L_{\rm H\beta}$ & \rb{Notes} \\
\noalign{\smallskip}\hline\noalign{\smallskip}
    1 & RX J0022--34 & 16.1 & 0.219 & 4110&\pl& 120 & 355&\pl& 10 
	& 60 & 3.5 & --0.40 & --0.17 & +0.23 & 36.1 \\
    2 & ESO 242--G8 & 16.1 &  0.059 & 3670&\pl& 160 & 310&\pl& 10 
	& 140 & 5.9 & --0.09 & +0.28 & +0.37 & 34.7 \\
    3 & WPVS 007 & 14.8 &     0.029 & 1620&\pl& 50 & 320&\pl& 40 
	& 215 & 3.6 & --0.82 & +0.68 & +1.50 & 34.5 \\ 
    4 & RX J0057--22 & 14.5 & 0.062 & 1380&\pl& 100 & 630&\pl& 160 
	& 185 & 3.3 & --0.90 & +0.72 & +1.62 & 35.1 \\
    5 & QSO 0056--36 & 15.1 & 0.165 & 4700&\pl& 160 & 435&\pl& 80 
	& 120 & 3.6 & --1.21 & +0.28 & +1.49 & 36.0  \\
    6 & RX J0100--51 & 15.4 &   0.062 & 3450&\pl& 120 & 565&\pl& 30 
	& 200 & 2.7 & --0.49 & +0.55 & +1.04 & 34.9 \\
    7 & MS 0117--28 & 16.0 &   0.349 & 2925&\pl& 100 & 895&\pl& 200 
	& 170 & 2.5 & --1.25 & +0.59 & +1.84 & 36.3 \\
    8 & IRAS 01267--21 & 15.4 & 0.093 & 2900&\pl& 100 & 410&\pl& 20 
	& 205 & 3.0 & --0.53 & +0.44 & +0.97 & 35.4 \\
    9 & RX J0134--42 & 16.0 &   0.237 & 1160&\pl& 80 & &  --- &  
	& 175 & 4.0 & --1.32 & +1.09 & +2.42 & 35.3 & 1 \\ 
   10 & RX J0136--35 & 18.0 &   0.289 & 1320&\pl& 120 & 870&\pl& 340 
	& 285 & 4.2 & --0.64 & +0.92 & +1.57 & 35.1 \\
   11 & RX J0148--27 & 15.5 &   0.121 & 1250&\pl& 100 & 765&\pl& 360 
	& 315 & 3.4 & --1.09 & +0.69 & +1.77 & 35.5 \\
   12 & RX J0152--23 & 15.6 &   0.113 & 3510&\pl& 130 & 680&\pl& 30 
	& 215 & 2.8 & --0.30 & +0.50 & +0.80 & 35.5 \\
   13 & RX J0204--51 & 16.6 &   0.151 & 5990&\pl& 240 & 310&\pl& 20 
	& 170 & 4.3 & --0.52 & +0.22 & +0.75 & 35.5 \\
   14 & RX J0228--40 & 15.2 &   0.494 & 2265&\pl& 100 & 700&\pl& 120
	& 215 & 4.0 & --1.30 & +0.64 & +1.94 & 37.0 \\
   15 & RX J0319--26 & 15.9 &   0.079 & 4170&\pl& 240 & 480&\pl& 40
	& 235 & 3.9 & --0.51 & +0.60 & +1.11 & 35.0 \\
   16 & RX J0323--49 & 16.5 &   0.071 & 2075&\pl& 70 &  230&\pl& 20
	& 205 & 4.4 & --0.66 & +0.52 & +1.18 & 34.7 \\
   17 & ESO 301--G13 & 15.5 &   0.064 & 3180&\pl& 200 & 545&\pl& 20
	& 195 & 2.9 &  +0.15 & +0.36 & +0.21 & 35.1 \\
   18 & VCV 0331--37 & 16.3 &   0.064 & 2165&\pl& 60 & 170 &\pl & 10
	& 205 & 3.2 & --0.59 & +0.39 & +0.98 & 34.8  \\
   19 & RX J0349--47 & 16.8 &   0.299 & 2250 &  \pl & 150 & 635&\pl& 60 
	& 530 & 3.5 & --0.50 & +0.64 & +1.14 & 36.0 \\
   20 & Fairall 1116 & 15.2 & 0.059 & 4560&\pl& 120 & 320&\pl& 20  
	& 130 & 3.5 & --0.67 & +0.20 & +0.88 & 35.2 \\
   21 & RX J0412--47 & 15.9 & 0.132 & 5580&\pl& 400 &  70&\pl& 10 
	& 45 & 5.8 & --0.17 & --0.09 & +0.08 & 35.5 & 2 \\
   22 & RX J0426--57 & 14.1 &   0.104 & 2900&\pl& 100 & 450&\pl& 10 
	& 30 & 4.2 &  +0.13 & --0.39 & --0.52 & 36.0 & 2 \\
   23 & Fairall 303 & 16.2 &   0.040 & 1720&\pl& 60 & 140&\pl& 10 
	& 370 & --- & --0.57 & +0.48 & +1.04 & 34.5 \\  
   24 & RX J0435--46 & 17.1 &   0.070 & 3820&\pl& 240 & 260&\pl& 20
	& 190 & 3.7 & --0.16 & +0.77 & +0.93 & 34.2 \\
   25 & RX J0435--36 & 17.1 &   0.141 & 6750&\pl& 620 & 810&\pl& 40 
	& 240 & 3.5 & --0.06 & +0.57 & +0.63 & 35.0 \\
   26 & RX J0437--47 & 15.3 &   0.052 & 4215&\pl& 120 & 240&\pl& 20
	& 130 & 3.2 & --0.71 & +0.38 & +1.08 & 34.9 \\
   27 & RX J0438--61 & 15.7 &   0.069 & 2410&\pl& 100 & 190&\pl& 10
	& 155 & 3.8 & --0.22 & +0.19 & +0.42 & 35.2 \\
   28 & RX J0439--45 & 16.6 &   0.224 & 2105&\pl& 100 & 1020&\pl& 240 
	& 210 & 3.5 & --0.77 & +0.65 & +1.42 & 35.6 \\
   29 & RX J0454--48 & 17.7 &   0.363 & 1970&\pl& 200 & &  --- & 
	& 265 & 4.4 &   ---  & +0.74 & --- & 35.6 & 3 \\
   30 & RX J1005+43 & 16.4 &   0.178 & 2990&\pl& 120 &  825&\pl& 150  
	& 330 & --- & --0.95 & +0.89 & +1.85 & 35.4 \\
   31 & CBS 126 & 15.4 &   0.079 & 2850&\pl& 200 &  370&\pl& 20 
	& 110 & 3.5 & --0.59 &  +0.13 & +0.72 & 35.3 \\
   32 & RX J1014+46 & 17.1 &   0.324 & & ---& & & --- & 
	& 230 & --- & --- & --- & --- & --- & 4 \\
   33 & RX J1017+29 & 15.7 &   0.049 & 1990&\pl& 200 & 255&\pl& 20 
	& 105 & --- & --0.16 & +0.65 & +0.81 & 34.3 \\
   34 & Mkn 141 & 15.1 &   0.042 & 4175&\pl& 340 &  395&\pl& 30 
	& 125 & 3.8 & --0.34 & +0.61 & +0.95 & 34.5 \\
   35 & Mkn 142 & 15.2 &   0.045 & 1790&\pl& 70 & 280&\pl& 30
	& 285 & 3.3 & --0.72 & +0.77 & +1.49 & 34.7 \\
   36 & RX J1050+55 & 16.7 &   0.333 &  & --- & & &  --- &  
	&  355 & --- & ---   & --- & --- & --- & 4 \\
   37 & EXO 1055+60 & 16.9 &   0.149 & 2155&\pl& 100 & 540&\pl& 50 
	& 365 & --- & --0.53 & +0.68 & 1.22  & 35.2 \\
   38 & RX J1117+65 & 16.4 &   0.147 & 2160&\pl& 110 & 880&\pl& 150
	& 260 & --- & --0.67 & +0.77 & +1.44 & 35.2 \\
   39 & Ton 1388 & 14.4 &   0.177 & 2920&\pl& 80 & 920&\pl& 150 
	& 190 & 4.6 & --1.04 & +0.50 & +1.54  & 36.3 \\
   40 & Mkn 734 & 14.4 &   0.033 & 2230&\pl& 140 & 450&\pl& 10 
	& 235 & 3.2 & --0.31 & +0.66 & +0.97 & 34.8 \\
   41 & Z 1136+34 & 16.0 &   0.033 & 1685&\pl& 80 & 210&\pl& 20 
	& 260 & --- & --0.57 & +0.56 & +1.13 & 34.3  \\
   42 & CSO 109 & 16.3 &   0.059 & 2270&\pl& 240 & 230&\pl& 20 
	& 185 & --- & --0.30 & +0.95 & +1.25 & 34.2 \\
   43 & RX J1231+70 & 16.0 &  0.208 & 3990&\pl& 150 &  550&\pl& 60 
	& 50 & --- & --0.41 & +0.11 & +0.52 & 35.7 & 2, 5 \\
   44 & IC 3599 & 16.5 &   0.021 &  635&\pl& 100 & 580&\pl& 30 
	&  65 & 4.2 & +0.50 &  +0.91 & +0.40 & 32.8 & 2 \\
   45 & IRAS 1239+33 & 15.1 &   0.044 & 1900&\pl& 150 & 485&\pl& 30 
	& 200 & 5.9 & +0.33 & +0.85 & +0.52 & 34.4 \\
   46 & RX J1312+26 & 16.2 &   0.061 & 2905&\pl& 220 & 205&\pl& 30 
	& 115 & --- & --0.82 & +0.60 & +1.42 & 34.4  \\
   47 & RX J1314+34 & 16.3 &   0.075 & 1590&\pl& 100 & 400&\pl& 50
	& 250 & --- & --0.58 & +0.71 & +1.29 & 34.7 \\
   48 & RX J1355+56 & 16.5 &   0.122 & 1780&\pl& 170 &  580&\pl& 20
	& 160 & --- &  +0.21 & +0.52 & +0.30 & 35.0 \\
   49 & RX J1413+70 & 16.9 &   0.107 & 5170&\pl& 400 & 395&\pl& 30 
	&  150 & --- & +0.08 & +0.60 & +0.52 & 34.7  \\
   50 & Mkn 684 & 14.7 &   0.046 & 1685&\pl& 100 & 405&\pl& 80 
	& 240 & 3.1 & --0.80 & +0.92 & +1.72 & 34.7 \\
   51 & Mkn 478 & 14.6 &   0.077 & 1915&\pl& 90 & 610&\pl& 110 
	& 290 & 3.7 & --0.78 & +0.72 & +1.50 & 35.5 \\
   52 & RX J1618+36 & 16.6 &   0.034 &  830&\pl& 80 & 100&\pl& 10 
	& 100 & --- & --0.22 & +0.77 & +0.99 & 33.5 \\
   53 & RX J1646+39 & 17.1 &   0.100 & 2160&\pl& 130 & 230&\pl& 20
	& 255 & --- & --0.50 & +0.62 & +1.12 & 34.7 \\
   54 & RX J2144--39 & 18.0 &  0.140 & 1445&\pl& 120 & 210&\pl& 20
	& 130 & 4.2 & --0.07 & +0.71 & +0.78  & 34.4 \\
   55 & RX J2154--44 & 15.8 &   0.344 & 2655&\pl& 100 & 510&\pl& 50 
	& 100 & 4.1 & --0.78 & +0.23 & +1.01 & 36.4 \\
\noalign{\smallskip}\hline\noalign{\smallskip}
\end{tabular}
\end{flushleft}
\end{table*}
\begin{table*}
\begin{flushleft}
\begin{tabular}{rlccr@{}l@{}rr@{}l@{}rccccccl}
\hline\noalign{\smallskip}
& & & & \multicolumn{6}{c}{FWHM} & EW & & log & log & log & log \\
\rb{No.} & \rb{Object} & \rb{V} & \rb{z} & 
\multicolumn{3}{c}{H$\beta$} & \multicolumn{3}{c}{[OIII]} & FeII &
\rb{$\rm \frac{H\alpha}{H\beta}$} & $\rm \frac{[OIII]}{H\beta}$ & 
$\rm \frac{FeII}{H\beta}$ & $\rm \frac{FeII}{[OIII]}$ 
& $L_{\rm H\beta}$ & \rb{notes} \\
\noalign{\smallskip}\hline\noalign{\smallskip}
   56 & RX J2213--17 & 17.2 &   0.146 & 1625 & \pl & 200 & 250 & \pl & 70 
	&  160 & 5.4 & +0.30 & +0.74 & +0.44 & 34.8 \\
   57 & RX J2216--44 & 15.8 &   0.136 & 2200 & \pl & 80 & 695 & \pl & 70 
	& 250 & 3.1 & --0.69 & +0.69 & +1.38 & 35.5 \\
   58 & RX J2217--59 & 16.2 &   0.160 & 1850 & \pl & 100 & 1075 & \pl & 150 
	& 330 & 3.4 & --0.51 & +0.99 & +1.49 & 35.2 \\
   59 & RX J2221--27 & 17.7 &  0.177 & 2090 & \pl & 100 & 465 & \pl & 40 
	& 265 & 3.5 & --0.52 & +0.53 & +1.04 & 35.3 \\
   60 & RX J2232--41 & 16.9 &  0.075 & 4490 & \pl & 350 & 455 & \pl & 20
	& 130 & 3.6 & --0.19 & +0.20 & +0.39 & 34.7 \\
   61 & RX J2241--44 & 15.8 &  0.545 & 1890 & \pl & 200 & 380 & \pl & 100
	& 210 & --- & --0.94 & +0.90 & +1.84 & 36.5 \\
   62 & RX J2242--38 & 16.9 &  0.221 & 2210 & \pl & 70 & 560 & \pl & 60 
	& 430 & 2.7 & --0.76 &  +0.62 & +1.38 & 35.7 \\
   63 & RX J2245--46 & 14.8 &  0.201 & 2760 & \pl & 140 & 680 & \pl & 60 
	& 205 & 3.4 & --0.46 & +0.52 & +0.97 & 36.4 \\
   64 & RX J2248--51 & 15.5 &  0.102 & 3460 & \pl & 200 & 230 & \pl & 10
	& 100 & 4.4 &  +0.02 & +0.23 & +0.21 & 35.5 \\
   65 & MS 2254--37 & 15.0 &   0.039 & 1545 & \pl & 70 & 610 & \pl & 50 
	& 90 & 4.4 & --0.44 &  +0.28 & +0.72 & 34.7 \\
   66 & RX J2258--26 & 16.1 &   0.076 & 2815 & \pl & 180 & 285 & \pl & 10 
	& 185 & 3.8 &  +0.15 & +0.36 & +0.22 & 35.0 \\
   67 & RX J2301--55 & 15.4 &   0.140 & 2590 & \pl & 120 & 355 & \pl & 80 
	& 230 & 3.7 & --1.08 & +0.71 & +1.78  & 35.6 \\
   68 & RX J2303--55 & 17.5 &  0.084 & 4030 & \pl & 200 & 495 & \pl & 60 
	& 160 & 3.9 & --0.75 & +0.71 & +1.46 & 34.2 \\
   69 & RX J2304--35 & 16.6 &  0.042 & 1775 & \pl & 130 & 210 & \pl & 10
	& 110 & 5.4 &  +0.39 & +0.66 & +0.27 & 34.0 \\
   70 & RX J2304--51 & 17.2 &  0.106 & 3830 & \pl & 160 & 190 & \pl & 10 
	& 85 & 3.1 & --0.05 & --0.16 & --0.11 & 35.0 \\
   71 & RX J2317--44 & 16.8 &  0.134 & 1390 & \pl & 35 &  350 & \pl & 50
	& 310 & 3.9 & --0.76 & +0.58 & +1.34 & 35.2 \\
   72 & RX J2325--32 & 17.0 &  0.216 & 3295 & \pl & 130 & 220 & \pl & 10
	& 145 & 4.6 & --0.43 & +0.01 & +0.44 & 35.8 \\
   73 & RX J2340--53 & 17.6 &  0.321 & 1880 & \pl & 120 &  710 & \pl & 90
	& 285 & 2.8 & --0.57 & +0.59 & +1.15 & 35.7 \\
   74 & MS 2340--15 & 15.6 &  0.137 & 1565 & \pl & 80 &  760 & \pl & 70
	& 175 & 3.5 & --0.86 & +0.61 & +1.47 & 35.5 \\
   75 & RX J2349--33 & 16.6 & 0.144 & 7715 & \pl & 850 & 560 & \pl & 10
	& 355 & 4.2 &  +0.42 & +0.11 & --0.30  & 35.9 & 2 \\
   76 & RX J2349--31 & 16.6 &  0.135 & 5210 & \pl & 270 & 475 & \pl & 20 
	& 95 & 3.0 & --0.15 &  +0.40 & +0.54  & 35.1 \\
\noalign{\smallskip}\hline
\end{tabular}
\end{flushleft}
Table notes: 
\begin{enumerate}
\vspace{-2mm}
\item In an optical spectrum of 5 hours exposure (Grupe et al., in
preparation), weak [OIII]$\lambda$5007 is visible.
\item Quantities referring to FeII are upper limits.
\item {[OIII]} is not in the observed wavelength range.
\item H$\beta$ and {[OIII]} are not in the observed wavelength range.
\item {[OIII]} measurements are based on the [OIII]$\lambda$4959 line.
\end{enumerate}
\end{table*}
}

All objects in our sample were found to be Seyfert 1 type AGN. We do
not divide the Seyfert 1 class into subclasses such as NLSy1s or
Seyfert 1.5\footnote{ The definitions underlying Seyfert
classifications depend on the resolution of and the noise in the
spectra used.  For instance, with increasing resolution, NLR emission
components will be found in Seyferts which defied such detection with
lower resolution, causing the definition of a certain object as
Seyfert 1.5 or NLSy1 to become fuzzy (see also Goodrich 1989a,b).  A
dividing line at 2000 {\kms} is generally used to separate Sy1 and
NLSy1, although many objects display NLSy1 properties in spite of
their FWHM(\hb) somewhat exceeding this value (e.g. Ton
1388). Furthermore, published values of FWHM(\hb) do not always refer
to the same component (ideally the broad one) of H$\beta$.  For these
reasons, we relax our language in the following refering to
Narrow-Line Seyfert 1 galaxies as Sy1 galaxies with narrow BLR
emission lines, weak [OIII]/\hb~ ratio, strong FeII emission, and
steep X-ray spectra when compared with canonical Sy1s.}. This soft
X-ray selected sample includes only one possible Sy 2 object, the
X-ray transient \mbox{IC 3599} which was discussed by Brandt et al. (1995) and 
Grupe et al. (1995a, 1998b). Recently, this object has been classified as a
Seyfert 1.9 by Komossa \& Bade (1999).

\begin{figure}[t]
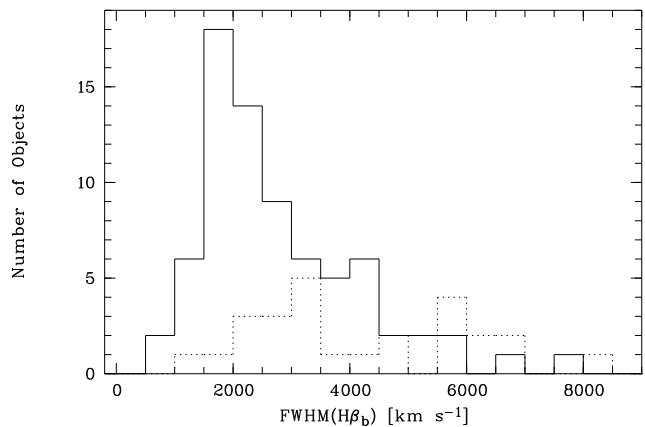

\clipfig{1029F001}{87}{12}{18}{280}{195}
\caption[ ]{\label{distr_fwhm} Distribution of the FWHM of \hb~ for
the soft X-ray selected AGN (solid line) and the hard the X-ray
comparison sample (dotted line)}
\end{figure}

The distribution of the H$\beta$ line width of the objects in our
sample of soft X-ray selected AGN shows a preference for rather narrow
permitted emission lines. The average line width is significantly
smaller than that of the objects from the comparison sample of hard
X-ray selected Seyfert galaxies. The distributions of the FWHM of the
broad components of H$\beta$ in both samples are displayed in
Fig.~\ref{distr_fwhm}. We find mean values of 2790$\pm$160 (median
2250) \kms and 4210$\pm$360 (median 3460) \kms, for the soft and hard
sample, respectively. Because some of the spectra of the hard X-ray
objects given in the literature are rather poor we may have
underestimated the widths of the broad components and the real
difference may be even larger. We again emphasize that our FWHM refer
to the broad component in H$\beta$ only, while many values quoted in
the literature refer to the whole line and may underestimate the FWHM
of the underlying broad component.

\subsection{ Correlations}

In this Sect. we show the results of correlation analyses performed
on parameters measured from the individual emission lines. The methods
used include the Principal Component Analysis (PCA, see Sect.
\ref{pca_sect} below) which is helpful in identifying the nature of
the physical parameters responsible for the observed correlations
between observed parameters.  We also present results of a Spearman
rank order test in Table~\ref{corr-test}. Correlations among the
continuum properties were already discussed in Paper I and are quoted
here only for comparison where needed.

\begin{table*}
\caption[Correlation coefficients of Spearman rank order test]
{\label{corr-test} Results of the the non-parametric Spearman rank
order test.  The lower left of the Table lists the correlation
coefficients and the upper right the significance levels obtained from
Student's t test, respectively.}
\begin{flushleft}
\begin{tabular}{lcccccccccc}
\hline \\
\noalign{\smallskip}
Prop & FWHM(\hb) & FWHM([OIII]) & $\rm \frac{[OIII]}{H\beta_{b}}$ & EW FeII &
\ax & \aopt & $\alpha_{ox-soft}$ & $\nu L_{\rm V}$ & $\nu L_{\rm 250eV}$ & 
$\rm \frac{H\alpha}{H\beta}$ \\
\noalign{\smallskip}\hline\noalign{\smallskip} 
FWHM(\hb) & --- & --0.5 & +2.0 & --2.8 & --3.3 & +0.3 & +1.5 & +1.1 & 
+0.3 & --0.6 \\ \noalign{\smallskip}
FWHM([OIII] & --0.06 & --- & --2.5 & +3.2 & +2.4 & --1.9 & --0.6 & +5.4 & 
+4.8 & --2.3 \\ \noalign{\smallskip}
{[OIII]}/{\hb} & +0.23 & --0.29 & --- & --3.3 & --2.6 & 
+6.0 & --1.1 & --3.8 & --3.2 & +2.2 \\ \noalign{\smallskip}
EW FeII  & --0.31  & +0.36 & --0.36 & --- & +1.9 & --1.1 & --0.9 & 
+1.3 & +1.6 & --2.2  \\ \noalign{\smallskip}
\ax & --0.36  & +0.27 & --0.29 & +0.22 & --- & --4.1 & --4.1 & +3.1 &
+5.8 & +0.3 \\ \noalign{\smallskip}
\aopt & +0.04  & --0.22 & +0.58 & --0.13 &  --0.43 & --- & 
+0.4 & --5.5 & --5.5 & +1.6 \\ \noalign{\smallskip}
 $\alpha_{ox-soft}$ & +0.17 & --0.07 & --0.13 & --0.10 &  --0.44 
& +0.05 & --- & +0.2 & --4.0 & --0.7 \\ \noalign{\smallskip}
$\nu L_{\rm V}$  & +0.13  & +0.54 & --0.41 & +0.15 &  +0.34 & --0.54 &  +0.02 
& --- & +15.2 & --0.8 \\ \noalign{\smallskip}
$\nu L_{\rm 250eV}$  & +0.04 & +0.50 & --0.35 &  +0.18 & +0.56 & --0.54 
& --0.42 & +0.87 & --- & --0.0 \\ \noalign{\smallskip}
$\rm \frac{H\alpha}{H\beta}$  & --0.08 & -0.29 & +0.27 & --0.27 & +0.04 
& +0.21 & --0.09 & --0.10 & --0.00 & --- \\ \noalign{\medskip}
\# of objects & 74 & 72 & 73 & 76 & 76 & 76 & 76 & 76 & 76 & 61 \\ 
\noalign{\smallskip}\hline\noalign{\smallskip}
\end{tabular}
\end{flushleft}
\end{table*}

\subsubsection{FWHM(\hb) -- \ax~ correlation}

\begin{figure}[t]
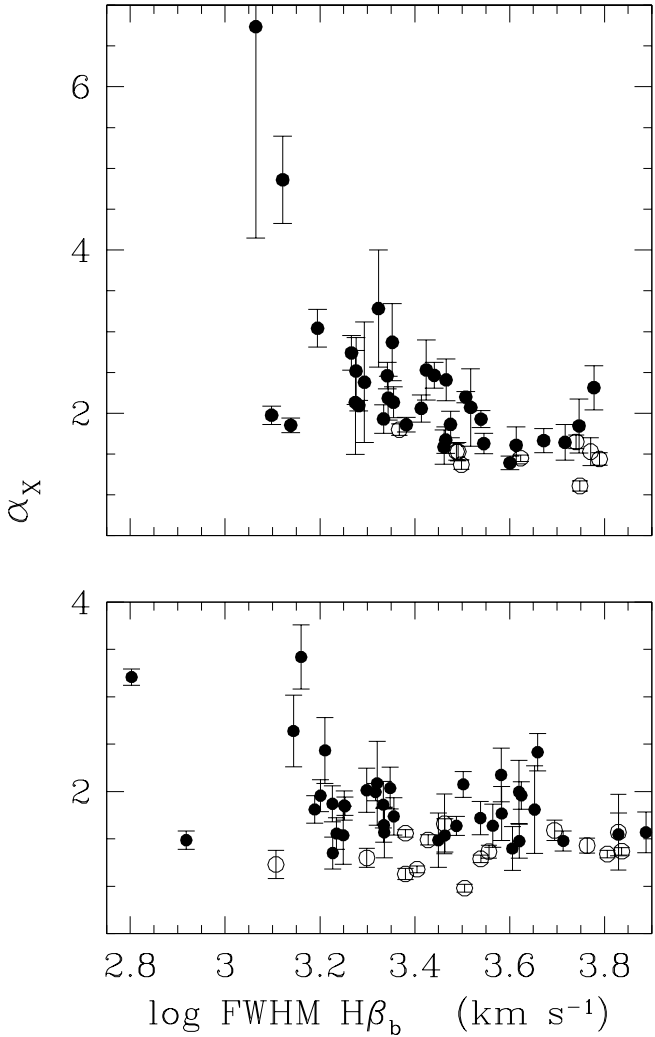

\clipfig{1029F002}{87}{17}{11}{180}{273}
\caption[ ]{\label{fwhb_ax} X-ray slope \ax{} vs. FWHM(\hb).  The
upper graph shows the high-luminosity AGN with log $\nu L_{\rm
V}>37.15$ Watt. The soft X-ray AGN are displayed as solid and the
hard X-ray AGN open circles.  The lower graph shows the low-luminosity
soft X-ray AGN with log $\nu L_{\rm V}<37.15$. The AGN with the
steepest X-ray spectrum, WPVS007, is off the plot (FWHM(\hb)=1620 $\rm
km~s^{-1}$, \ax=8.0)  }
\end{figure}

An anti-correlation in the line width of \hb~ and the X-ray slope
$\alpha_x$ is known from studies of optically selected NLSy1 galaxies
(e.g., Boller et al. 1996) and quasars (Laor et al.~1997) and is
prominent also in our sample of soft X-ray selected AGN
(Fig.~\ref{fwhb_ax}). We find that only objects with comparatively
narrow \hb~ emission lines have steep X-ray spectra. Objects with
broad \hb{} lines always have harder X-ray spectra.  A similar result
was found by Boller et al. (1996). A Spearman rank order test yields a
correlation coefficient of $r_s = -0.36$, corresponding to a
significance level of Student's t-test of $t = -3.3$ or a probability
of 0.001 of finding this result in random data (see
Table~\ref{corr-test}).  Dividing our sample into two subsamples with
the dividing line at the average X-ray luminosity, log~$\nu L_V~=$
37.15 (in Watts, see Paper I), we find that the correlation is more pronounced
among the AGN with higher luminosity where the significance becomes
$r_s = 0.59$ ($t = -4.1$) for the soft X-ray selected AGN alone and
$r_s = -0.71$ ($t = -6.4$) for the hard and soft X-ray selected AGN
together. The latter result is reminiscent of that found by Laor et
al. (1997).  The correlation is much weaker or absent among the
low-luminosity objects. Whatever causes the differences in \ax{} in
the ROSAT bandpass, it must be a property of the AGN itself rather
than of the host galaxies, since FWHM(\hb) and \ax~ both refer to
nuclear properties. The object with the steepest X-ray spectrum,
WPVS007 (FWHM(\hb)=1620 km\,s$^{-1}$ and \ax{} = 8.0, Grupe et al. 1995b)
is off scale in Fig.~\ref{fwhb_ax}, but is included in the correlation
analysis.

\subsubsection{FWHM([OIII]) -- luminosity correlation}

\begin{figure}[t]
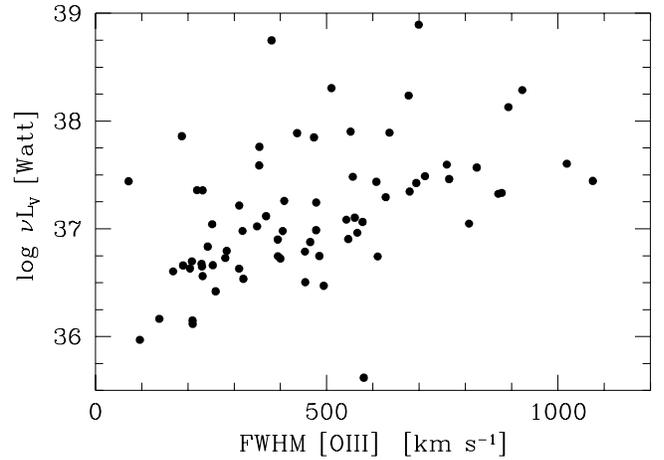

\parbox[b]{9cm}{
\clipfig{1029F003}{87}{13}{63}{195}{190}
}
\caption[ ]{\label{fwo3_lv}
FWHM [OIII] vs. optical monochromatic luminosity log$\nu L_{\rm V}$
}
\end{figure}

The correlations between the FWHM([OIII]) and the optical luminosity
as well as the X-ray luminosity at 250 eV are significant at the
$t\sim$ 5 level.  Fig. \ref{fwo3_lv} displays the correlation
between the [OIII] line width and the optical luminosity.

\subsubsection{Correlations involving [OIII]/\hb}

\begin{figure}[t]
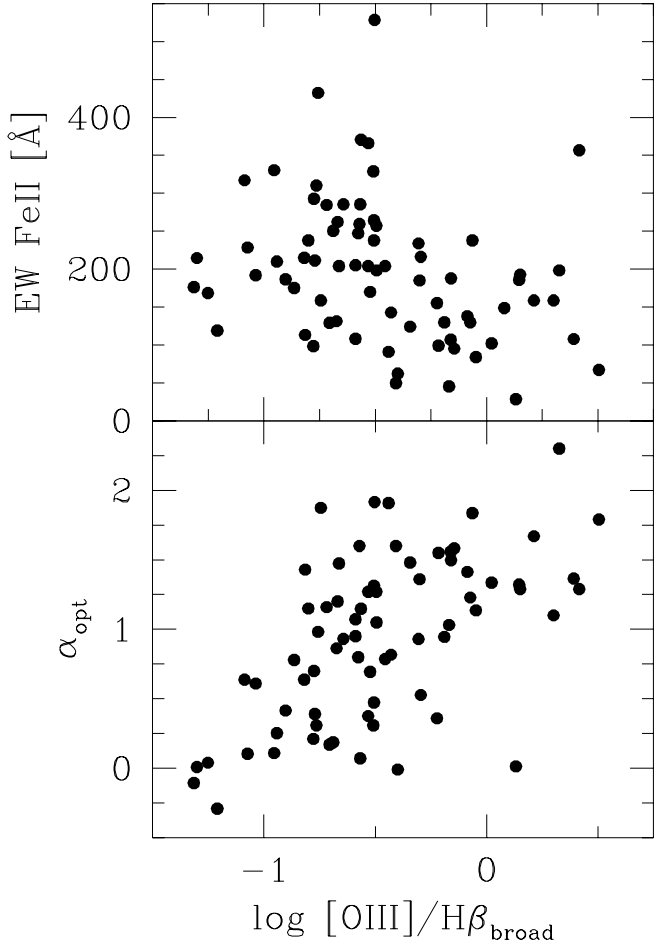

\parbox[b]{9cm}{
\clipfig{1029F004}{87}{41}{61}{162}{240}
}
\caption[ ]{\label{o3hb_ewfe2_aopt}
[OIII]/H$\beta_{\rm broad}$ vs. EW FeII and \aopt. A small value of 
\aopt~ means a blue coninuum and a high value represents a steeper, redder
optical spectrum
}
\end{figure}

We find an anti-correlation between the flux ratio [OIII]/\hb~ and the
equivalent width of FeII with a correlation coefficient $r_{\rm s} =
-0.36$ ($t = -3.3$). Such correlation is already known from other
studies (e.g.  Boroson \& Green 1992, or Laor et al. 1997). There is
also a strong correlation between [OIII]/\hb~ and the optical
continuum slope \aopt{} with $r_{\rm s} = 0.58$ ($t = 6.0$).  Fig.
\ref{o3hb_ewfe2_aopt} shows both results.

\subsubsection{Internal Absorption}

In order to check which effect internal absorption in the AGN may have
on some of our results, we searched for correlations (i) between the
optical continuum slope and the Balmer decrement H$\alpha$/H$\beta$
and (ii) between the column density of cold absorbing matter as
derived from the X-ray spectra and the Balmer decrement. None was
found, suggesting that internal absorption does not seriously affect
our results. This is demonstrated in Fig. \ref{hahb_nh_aopt} and Table
\ref{corr-test} where the quantity $\rm \Delta N_H$ represents the
difference between the hydrogen column density determined from an
absorbed power-law fit to the X-ray spectrum and the total galactic
column density of atomic hydrogen (see Paper I). The mean of the
measured Balmer decrements is H$\alpha$/H$\beta$ = 3.80$\pm$0.10
(median 3.66) which is not far from the case B value 3.10
(e.g. Osterbrock 1989). Although case B may not be applicable in our
case, the majority of the soft X-ray selected AGN is not substantially
reddened. This is also consistent with the finding that only the
object with the highest Balmer decrement, IRAS F12397+3333, is
polarized, while all other members of the sample show are
marginally polarized or not at all (Grupe et al. 1998b). 

\begin{figure}[t]
\parbox[b]{9cm}{
\clipfig{1029F005}{87}{43}{62}{162}{240}
}
\caption[ ]{\label{hahb_nh_aopt}
Balmer decrement H$\alpha$/H$\beta$ vs. $\Delta N_{\rm H}$ and \aopt
}
\end{figure}

\subsection{\label{pca_sect} Principal Component Analysis}

The appearance of AGN is governed by several underlying physical
parameters. Correlation analyses between two observational quantities
can reveal only part of the picture. Principal Component Analysis
(PCA) is a method by which information on the underlying physical
parameters can be obtained. It is based on a transformation to the
principal axes of the ellipsoid formed by the data points in
$n$-dimendional space. PCA reduces the number of relevant components
and these remaining components are considered more basic than the
observed properties and more astrophysically meaningful such as, in
fortunate circumstances, the mass of the central black hole or the
accretion rate.  PCA was originally used in social studies but has
been successfully applied in astrophysics as well (see e.g. Whitney
1983a,b, Boroson \& Green 1992, Francis et al. 1992, Brotherton 1996).

The PCA was applied to 72 objects using the following quantities: the
FWHM of \hb~and [OIII], the flux ratio [OIII]/\hb, the equivalent
width of FeII, the X-ray spectral index \ax, the spectral index
\aopt of the optical continuum, and log $\nu L_{\rm V}$. The results
are presented in Table \ref{pca}. We find that the first principal
component accounts for 40\% of the intrinsic variance of the input
parameters and, hence, seems to be an important underlying parameter
that governs the observed properties of the AGN. Also the second and
third components are still strong. The first three components together
account for more than 70\% of the variance.  The first component has a
strong effect on FWHM([OIII]), [OIII]/\hb, \aopt, and the luminosity
(see Table \ref{pca}).  It does not, however, show the correlation
between \ax~ and FWHM(\hb).  This result differs from that of Boroson
\& Green (1992) and Laor et al. (1997) who found that this correlation
affects their Eigenvector 1. We have shown that the \ax-FWHM(\hb)
correlation is present for the high-luminosity AGN but does not appear
in the low-luminosity fraction of our sample (see Fig.
\ref{fwhb_ax}) and, by introducing subsamples with different cuts in
luminosity, we confirmed that this is indeed the reason for the
difference to the Boroson \& Green and Laor et al. results.

\section{\label{dis} Discussion}

Soft X-ray selection of bright AGN brings out mainly so-called
Narrow-Line Seyfert 1 galaxies.  Given the
selection bias in favor of low intrinsic column density objects,
type-1 Seyferts were strongly expected from the unified model for
Seyfert galaxies in which type-2 Seyferts are seen through large
absorbing column depths associated with molecular tori surrounding the
central accretion black hole and therefore do not show up as strong
soft X-ray sources.  Among the newly found ROSAT AGN, there is a
peculiar trend with luminosity.  The most luminos sources in the
sample have the bluest optical continua, the steepest X-ray spectra,
the broadest [OIII] but narrower {\hb} line widths, the weakest
[OIII]/{\hb} flux ratios, and the strongest optical FeII blends.  In
Paper I we argued that these sources also seem to have the most
pronounced Big Blue Bump spectra extending from optical wavelengths to
soft X-rays peaking somewhere in the EUV\footnote{ This trend does not
seem to continue far into the quasar luminosity range.  Quasar spectra
bend sharply in the UV to meet the X-ray power laws (Zheng et
al. 1997).  Such a spectral shape cannot be accepted for the AGN in
this study, since the extrapolation of the steep X-ray slopes toward
the UV would imply unphysical luminosities in the EUV.}.  We recall
that some of the AGN in the sample also showed extreme X-ray
variability and spectral changes (see also \ref{transient}).
The drawback with optical emission
line studies is that they are sensitive to a very large number of
physical parameters such as the kinematics, the density and
temperature, and the composition of the emitting clouds, the ionizing
continuum from the active nucleus and the properties of the ambient
medium.  Certainly, a new class of AGN could differ in all aspects
from other well-known Seyfert galaxies, but in a more economic scheme
one would like to see just one fundamental quantity to be responsible
for the observed pecularities.  Among such quantities, those related
to the accreting black hole paradigm (black hole mass, mass accretion
rate, inclination angle, density of ambient medium, power of nuclear
wind) seem to be the most promising in holding the clue to the
explanation of the trends seen among the bright soft X-ray selected
AGN.

\begin{table}
\caption[ ]
{\label{pca} Result of the principal component analysis (PCA) for 72 AGN of the
soft X-ray selected ROSAT AGN sample}
\begin{flushleft}
\begin{tabular}{lccc}
\hline 
\noalign{\smallskip}
Property & E-vector 1 & E-vector 2 & E-vector 3 \\
\noalign{\smallskip}\hline\noalign{\smallskip} 
Eigenvalue & 2.772  & 1.407  & 0.933  \\
 Proportion & 0.396  & 0.201  & 0.133 \\
Cumulative & 0.396  & 0.597  & 0.730  \\
\noalign{\medskip}\hline\noalign{\medskip} 
FWHM(\hb) & --0.145 & +0.710 & --0.187   \\       
FWHM([OIII])   & +0.389 & +0.073  & --0.638 \\    
{[OIII]/\hb} & --0.439 & +0.048 & --0.145  \\     
EW FeII       & +0.312 & --0.435 & --0.452 \\     
\ax           & +0.375 & --0.200 & +0.381   \\    
\aopt         & --0.452 & --0.243 & --0.425  \\   
log $\nu L_{\rm V}$  & +0.438 & +0.447 & --0.083\\
\noalign{\smallskip}\hline\noalign{\smallskip}
\end{tabular}
\end{flushleft}
\end{table}

\subsection{FWHM(\hb) -- \ax~ correlation}

One of the most prominent relations found in X-ray-selected AGN
samples is the trend for objects with steeper X-ray spectra to show
narrower BLR lines. This does not necessarily hold the other way
around: objects with narrow BLR lines may show both steep as well as
flat X-ray spectra, while objects with broad emission lines always
have flat X-ray spectra. This upper right part of Fig. \ref{fwhb_ax}
is, therefore, sometimes referred to as the 'zone of avoidance'.
There are several possible explanations for this white spot on the
map: a large mass of the central black hole, low inclination, or
obscuration effects.

A simple explanation of the special nature of NLSy1 galaxies is a
lower mass of the black hole compared to the black hole mass of other
Sy1s. Keplerian rotation of the BLR clouds would then proceed at
comparatively small velocities and \hb~would be narrow. Lower-mass
black holes produce soft X-ray spectra for two reasons. First, due to
the $T~\propto~(\dot{M}/M)^{1/4}$ relation (e.g Laor \& Netzer 1989,
Ross et al.  1992), such black holes would result in a hotter
accretion disc, and second, in order to reach a high luminosity it
would have to accrete closer to its Eddington limit, causing the X-ray
spectrum to become steep (e.g. Ross et al. 1992).

Another explanation of the \ax~$-$FWHM(\hb) relation can be that the
size of the BLR is different in NLSy1s compared to 'normal' Seyfert
1s. Wandel (1997) and Wandel \& Boller (1998) suggested that the size
of the BLR and therefore the distance of the BLR clouds to the center
is directly related to the shape of the X-ray spectrum. In their
model, the occurrence of steep X-ray spectra increases with
luminosity. Low-luminosity NLSy1s would be those with flat X-ray
spectra.  This is in agreement with our finding of no correlation
between \ax~ and \hb~ for the low-luminosity AGN.

Orientation effects can also play a role to explain the
\ax$-$FWHM(\hb) relation. Wills \& Browne (1986) showed that the BLR
seems to be flattened. The X-rays are thought to be emitted from the
inner part of an accretion disk. If this disk is thicker in the inner
part, looking at low inclination angles would lead to narrow emission
lines from the BLR and higher X-ray fluxes at soft X-ray energies. At
higher inclination angles, we would see broader lines and flatter
X-ray spectra.

Related to orientation effects is also the possibility that the
high-velocity clouds in NLSy1 may be obscured. Our measurement of the
Balmer decrements does not support this view nor is there significant
polarization indicative of scattering in an obscuring medium (Grupe et
al. 1998b).

In summary, the observed \ax$-$FWHM(\hb) relation is consistent with
several simple schemes. The fact, however, that the narrow H$\beta$
lines typically go along with strong FeII emission line strength, low
[OIII]/\hb, and steep X-ray spectra calls for a more involved
explanation.

\subsection{Other correlations}

The Principal Component Analysis shows that the common trend among the
correlations between the various emission line and continuum
properties is indeed dominated by the variation of one fundamental
quantity, as can be seen from the dominance of Eigenvector 1 (EV1).
This corroborates results from from differently selected NLSy1s using
the same statistical analysis technique (Boroson \& Green 1992, Laor
et al. 1997).  This parameter is not the inclination angle (Lawrence
et al. 1997) which was already rejected based on the missing
polarization, lack of internal soft X-ray absorption, and the
uncorrelated Balmer decrements, since the [OIII] flux (believed to be
a robust indicator of the isotropic flux) should then be independent
of EV1 contrary to the result of the analysis.  In principle, the
accretion rate in units of the Eddington accretion rate works well to
drive the correlations by increasing the Big Blue Bump strength and by
decreasing the Balmer line width, but some additional element
connected with the higher accretion rate is missing in order to
explain the other contributions to EV1.  This missing element could be
the density of the ambient medium, the power of a nuclear wind, or the
duration of the near-Eddington episodes of the accretion mode.
Near-Eddington accretion could require that the central gas density is
higher.  The higher electron density could drive the correlation
between luminosity and FWHM(OIII) (Fig.3) provided that the ionizing
luminosity averaged over the light-crossing time of the NLR ($\sim
100$~years) is much lower than the observed luminosity (to account for
the weakness of the NLR). This means that one would have to require
that the high-states are of a transient nature (as observed in a few
individual cases).  The other properties that go with Eigenvector 1
then follow naturally from the enhanced accretion rate. The scenario
also accounts for the minor effect on FWHM(\hb), since the increase in
electron density is compensated by the increase in luminosity in the
vicinity of the accretion disk (i.e. if $t_{\rm dur}\ge t_{\rm
BLR}\sim $~months).  The higher density could also be a secondary
effect associated with the higher accretion rate in NLSy1s.  Lawrence
et al. (1997) suggested that EV1 could reflect the effect of the
density of an out-flowing wind. They argue that 'the low-excitation of
at least the BLR results from gas that is mechanically heated rather
than radiatively heated'. There are several observations that seem to
support this picture. First of all, outflows in AGN are observed in an
increasing number of sources. Speaking for this is the finding of UV
absorption lines in WPVS007 and RX~J0134.2--4258 (Goodrich et al. 1999, in
prep), the AGN with the steepest
X-ray spectra during the RASS. Recently it was also reported that the
Eigenvector 1 found in the PG sample of Laor et al. (1997) is rather
strongly related to a unique density indicator, (SiIII]/CIII])
(B.J. Wills, private communication). Similarly, Smith (1993) argued
for a nuclear wind that pushes the lighter clouds outwards. These can
form the NLR, while heavier clouds would fall inwards. In the profiles
of the [OIII] line we see a blue-shifted base in the weaker lines
which may be related to these suggestions. However, blending of
[OIII]$\lambda$5007 underlying FeII is important and contamination by
some not so well understood component of the Fe complex can not be
ruled out.

Alternatively, we may speculate that Eigenvector 1 represents the age of an AGN.
It is likely that the appearance as an AGN is a
short-lived phenomenon that can occur in every galaxy, rather than a long-lived
phenomenon in only a few systems (Bechtold et al. 1994). If we assume that
Eigenvector 1 is the age this would result in a higher accretion rate and
probably stronger out-flowing wind at the beginning of the AGN's life, implying
that our soft X-ray selected AGN are relatively young. The age would nicely 
connect the properties that can be explained with the accretion rate with those 
that seem to be connected with outflows.

\subsection{\label{transient} Transient Soft X-ray AGN}

Soft X-ray AGN and in particular NLSy1s often appear to by rapidly variable in
X-rays (see e.g. Boller et al. 1993, 1997, and Leighly 1999).
It should be noted that three of the sources of our bright soft X-ray AGN sample
appear to be transient in X-rays. While IC 3599 has shown an outburst in X-rays
with a response in its optical lines (Brandt et al. 1995. Grupe et al. 1995a),
WPVS007, the AGN with the softest X-ray spectrum seen during the RASS was 
practically off in a pointed observation three years later
(Grupe et al. 1995b). The third source, RX
J0134.2--4258, is somewhat different. While its count rate in the ROSAT band
remained nearly constant its X-ray spectrum changed completely between 
its RASS and a pointed observation two years later (Mannheim et al. 1996; 
Grupe et al. 1999). We found that it is the hard
component that is mostly responsible for the peculiar spectral behavior of 
this AGN.
In hard X-ray surveys this source would appear to be transient as well. 
All the other sources of
our sample do not show significant variations either in X-rays or it the
optical spectral range. 

\section{Summary and conclusions}
We have studied the emission line properties of a sample of 76 bright
soft X-ray-selected AGN and found that

\begin{itemize}
\item they are Seyfert 1s preferably with relatively narrow BLR
emission lines,

\item there is a general trend that the steep X-ray spectra also go
along with blue optical continua, weak forbidden emission lines,
strong optical FeII blends, and high luminosities,

\item this trend seems to be driven by one or very few physical
properties (Eigenvector 1 in a principal component analysis accounts
for 40\% of the intrinsic variance of the data),

\item the accretion rate in units of the Eddington accretion rate most
certainly is one basic driver behind Eigenvector 1,

\item short-lived states (months to years) of a higher than usual
accretion rate and higher electron densities of the emission line
clouds possibly associated with nuclear winds could play a role for
the observed trends.

\end{itemize}

\acknowledgements{We thank Dr. Klaus Reinsch for taking part of the
optical spectra, Dr.~Beverley Wills for the providing the 
Boroson \& Green FeII
template and for numerous helpful comments, and Drs. Karen Leighly,
Stefanie Komossa, and Thomas Boller for useful suggestions and
discussions. Special thanks go also to David Doss, Jerry Martin, Marian
Frueh, and Doug Otoupal at McDonald Observatory for their instrumental
and observing help.  This research has made use of the NASA/IPAC
Extragalactic Database (NED) which is operated by the Jet Propulsion
Laboratory, Caltech, under contract with the National Aeronautics and
Space Administration. Also we used the IRAS data request of the
Infrared Processing and Analysis Center (IPAC) Caltech. This research
was first supported by the DARA under grant 50 OR 92 10 and later
sponsored by the Bundesanstalt f\"ur Arbeit.  The ROSAT project is
supported by the Bundesministerium f\"ur Bildung und 
Forschung  (BMBF) and the Max-Planck-Gesellschaft.

This paper can be retrieved via WWW from our pre-print server: \\
http://eden.uni-sw.gwdg.de/preprints/preprints.html and
http://www.xray.mpe.mpg.de/~dgrupe/research/refereed.html
}

\appendix
\section{\label{opt-spec} Optical Spectra}

The instrument mounted at the telescope was the ESO Faint Object
Spectrograph and Camera (EFOSC2) fitted out with the following grisms:
\begin{itemize}
\item \#1: 3400 - 9200 \AA; 8.4 \AA/pix $\approx$~ 22~ \AA~ FWHM resolution 
(1.5'' slit)
\item \#4: 4650 - 6800 \AA; 2.2 \AA/pix $\approx$~ 7 ~\AA~ FWHM resolution
\item \#8: 4640 - 5950 \AA; 1.3 \AA/pix $\approx$~ 5~ \AA~ FWHM resolution
\item \#9: 5875 - 7020 \AA; 1.1 \AA/pix $\approx$~ 5~ \AA~ FWHM resolution
\item \#10: 6600 - 7820 \AA; 1.2 \AA/pix $\approx$~ 5~ \AA~ FWHM resolution
\end{itemize}
The EFOSC CCD camera was equipped with a THX 1024$\times$1024 CCD with
19 $\mu$m squared pixels.

The observations of all northern AGN were performed with the 2.1m Otto
Struve telescope at the McDonald Observatory/West-Texas (MCD2.1 in
Table \ref{opt-list}), which is run by the University of Texas at
Austin. For these observations, the Cassegrain Spectrograph ES2 was
mounted at the telescope.  During the March 1994 observing run,
grating \#22 (114 \AA/mm $\approx$~ 4~ \AA~FWHM resolution) was used
to examine the wavelength range between 4650 to 6050 \AA. The camera
was equipped with the CC1 CCD (1024$\times$1024; 12$\rm \times 12~ \mu
m^2$ pixel size).

{\footnotesize
\begin{table*}
\caption[Optical observations]{\label{opt-list}
List of the optical observations of the objects of the soft X-ray AGN
sample. ESO2.2 = ESO/MPI 2.2m telescope, La Silla; MCD2.1 = 2.1m
telescope McDonald Observatory, Texas; EFOSC = ESO Faint Object
Spectrograph and Camera, ES2 = McDonald Cassegrain Spectrograph. The
numbers given behind the instrument indicate the used grisms or
gratings.  The exposure time $\rm T_{exp}$ is given in minutes.  }
\begin{flushleft}
\begin{tabular}{rlccrcll}
\hline\noalign{\smallskip}
No. & Object & $\alpha_{2000}$ & $\delta_{2000}$ &
obs. date & Tel. & Instrument & $\rm T_{exp}$  \\
\noalign{\smallskip}\hline\noalign{\smallskip}
1 & RX J0022-34 & 00 22 33 & --34 07 22 &
93/10/11 & ESO2.2 & EFOSC 1,8,10 & 10,45,45    \\
2 & ESO242--G8 & 00 25 01 & --45 29 55 &
93/09/14 & ESO2.2 & EFOSC, 1,8,10 & 5,30,30    \\
3 & WPVS007 & 00 39 16 & --51 17 03 &
93/01/01 & ESO2.2 & EFOSC, 1,8,10 & 5,20,20   \\
4 & RX J0057--22 & 00 57 20 & --22 22 56 &
92/10/17 & ESO2.2 & EFOSC, 1,8,10 & 5,30,30   \\
5 & QSO0056--36 & 00 58 37 & --36 06 06 &
93/10/12 & ESO2.2 & EFOSC 1,8,10 & 10,25,20   \\
6 & RX J0100--51 & 01 00 27 & --51 13 55 & 
93/10/12 & ESO2.2 & EFOSC, 1,8,10 & 10,40,35   \\
7 & MS0117--28 & 01 19 36 & --28 21 31 & 
93/10/14 & ESO2.2 & EFOSC, 1,9 & 5,30   \\
8 & IRAS01267--2157 & 01 29 11 & --21 41 57 & 
93/12/18 & ESO2.2 & EFOSC 1,8,10 & 5,30,20   \\
9 & RX J0134--42 & 01 34 17 & --42 58 27 &
93/12/18 & ESO2.2 & EFOSC, 1,8,9 & 5,30,30   \\
10 & RX J0136--31 & 01 36 54 & --35 09 53 &
93/12/18 & ESO2.2 & EFOSC, 1,4 & 10,30   \\
11 & RX J0148--27 & 01 48 22 & --27 58 26 &
92/08/26 & ESO2.2 &EFOSC, 8,10 & 15,15    \\
12 & RX J0152--23 & 01 52 27 & --23 19 54 &
93/09/12 & ESO2.2 & EFOSC, 1,8,10 & 5,30,30   \\
13 & RX J0204--51 & 02 04 03 & --51 04 57 &
93/10/14 & ESO2.2 & EFOSC, 1,4 & 10,30   \\
14 & RX J0228--40 & 02 28 15 & --40 57 16 &
92/10/18 & ESO2.2 & EFOSC, 1,10 & 5,30   \\
15 & RX J0319--26 & 03 19 49 & --26 27 13 &
92/10/18 & ESO2.2 & EFOSC, 1,8,10 & 5,30,30   \\
16 & RX J0323--49 & 03 23 15 & --49 31 15 &
93/08/21 & ESO2.2 & EFOSC, 1,8,10 & 30,30,30 \\
17 & ESO301--G13 & 03 25 02 & --41 54 18 &
93/09/13 & ESO2.2 & EFOSC 1,8,10 & 5,45,45   \\
18 & VCV0331--37 & 03 33 40 & --37 06 55 &
93/09/14 & ESO2.2 & EFOSC 1,8,10 & 5,30,30   \\
19 & RX J0349--47 & 03 49 08 & --47 11 05 &
93/10/11 & ESO2.2 & EFOSC 1,9 & 10,40   \\
20 & Fairall 1116 & 03 51 42 & --40 28 00 &
93/10/12 & ESO2.2 & EFOSC, 1,8,10 & 5,30,20   \\
21 & RX J0412--47 & 04 12 41 & --41 12 46 &
93/10/19 & ESO2.2 & EFOSC, 1,8,10 & 5,30,30   \\
22 & RX J0426--57 & 04 26 01 & --57 12 02 & 
93/09/14 & ESO2.2 & EFOSC, 1,8,10 & 5,20,15   \\
23 & Fairall 303 & 04 30 40 & --53 36 56 &
93/10/10 & ESO2.2 & EFOSC 1,8,10 & 5,30,20   \\
24 & RX J0435--46 & 04 35 14 & --46 15 33 &
93/12/18 & ESO2.2 & EFOSC, 1,4 & 5,30   \\
25 & RX J0435-36 & 04 35 54 & --36 36 41 &
93/10/14 & ESO2.2 & EFOSC 1,4 & 5,30   \\
26 & RX J0437--47 & 04 37 28 & --47 11 29 &
92/08/26 & ESO2.2 & EFOSC 8,10 & 15,15   \\
27 & RX J0438--61 & 04 38 29 & --61 47 59 &
93/10/11 & ESO2.2 & EFOSC, 1,8,10 & 5,15,20   \\ 
28 & RX J0439--45 & 04 39 45 & --45 40 42 &
93/10/12 & ESO2.2 & EFOSC, 1,8,9 & 10,40,35   \\
29 & RX J0454--48 & 04 54 43 & --48 13 20 &
93/10/14 & ESO2.2 & EFOSC, 1,4 & 10,30   \\
30 & RX J1005+43 & 10 05 42 & +43 32 41 &
94/03/07 & MCD2.1 & ES2, 22 & 45   \\
31 & CBS 126 & 10 13 03 & +35 51 24 &
94/03/06 & MCD2.1 & ES2, 22 & 45  \\
32 & RX J1014+46 & 10 14 02 & +46 19 54 &
94/03/08 & MCD2.1 & ES2, 22 & 45   \\
33 & RX J1017+29 & 10 17 18 & +29 14 34 &
94/03/07 & MCD2.1 & ES2, 22 & 35   \\
34 & Mkn 141 & 10 19 13 & +63 58 03 &
94/03/06 & MCD2.1 & ES2, 22 & 30   \\
35 & Mkn 142 & 10 25 31 & +51 40 35 &
94/03/07 & MCD2.1 & ES2, 22 & 40   \\
36 & RX J1050+55 & 10 50 55 & +55 27 23 &
94/03/08 & MCD2.1 & ES2, 22 & 45   \\
37 & EXO1055+60 & 10 58 30 & +60 16 01 &
94/03/08 & MCD2.1 & ES2, 22  & 45   \\ 
38 & RX J1117+65 & 11 17 10 & +65 22 07 &
94/03/08 & MCD2.1 & ES2, 22 & 45   \\
39 & Ton 1388 & 11 19 09 & +21 19 18 &
94/03/06 & MCD2.1 & ES2, 22 & 45    \\
40 & Mkn 734 & 11 21 47 & +11 44 19 &
94/03/07 & MCD2.1 & ES2, 22 & 30   \\
\noalign{\smallskip}\hline\noalign{\smallskip}
\end{tabular}
\end{flushleft}
\end{table*}
}
\begin{table*}
\begin{flushleft}
\begin{tabular}{rlccrcll}
\hline\noalign{\smallskip}
No. & Object & $\alpha_{2000}$ & $\delta_{2000}$ &
obs. date & Tel. & Instrument & $\rm T_{exp}$  \\
\noalign{\smallskip}\hline\noalign{\smallskip}
41 & Z 1136+34 & 11 39 14 & +33 55 52 &
94/03/07 & MCD2.1 & ES2, 22 & 35   \\
42 & CSO 109 & 11 45 10 & +30 47 17 &
94/03/08 & MCD2.1 & ES2, 22 & 45   \\
43 & RX J1231+70 & 12 31 37 & +70 44 14 &
94/03/08 & MCD2.1 & ES2, 22 & 30   \\
44 & IC 3599 & 12 37 41 & +26 42 28 &
94/03/06 & MCD2.1 & ES2, 22 & 45    \\
45 & IRASF1239+33 & 12 42 11 & +33 17 03 &
94/03/07 & MCD2.1 & ES2, 22 & 40   \\
46 & RX J1312+26 & 13 12 59 & +26 28 27 &
94/03/07 & MCD2.1 & ES2, 22 & 45   \\
47 & RX J1314+34 & 13 14 23 & +34 29 40 &
94/03/06 & MCD2.1 & ES2, 22 & 45   \\
48 & RX J1355+56 & 13 55 17 & +56 12 45 &
94/03/08 & MCD2.1 & ES2, 22 & 45   \\ 
49 & RX J1413+70 & 14 13 37 & +70 29 51 &
94/03/06 & MCD2.1 & ES2, 22 & 35   \\
50 & Mkn 684 & 14 31 05 & +28 17 15 &
94/03/07 & MCD2.1 & ES2, 22 & 30   \\ 
51 & Mkn 478 & 14 42 08 & +35 26 23 &
94/03/06 & MCD2.1 & ES2, 22 & 45   \\
52 & RX J1618+36 & 16 18 09 & +36 19 58 & 
94/03/08 & MCD2.1 & ES2, 22  & 45   \\
53 & RX J1646+39 & 16 46 26 & +39 29 33 &
94/03/08 & MCD2.1 & ES2, 22 & 30   \\
54 & RX J2144 --39 & 21 44 06 & --39 49 01 &
92/08/21 & ESO2.2 & EFOSC, 1,8,10 
&  30,30,30   \\
55 & RX J2154--44 & 21 54 51 & --44 14 06 &
92/10/18 & ESO2.2 & EFOSC, 1,9 & 5,30   \\ 
56 & RX J2213--17 & 22 13 00 & --17 10 18 &
93/08/20 & ESO2.2 & EFOSC, 1,8,10 
& 10,30,15    \\
57 & RX J2216--44 & 22 16 53 & --44 51 57 &
92/10/17 & ESO2.2 & EFOSC, 1,8,10 & 5,30,30   \\
58 & RX J2217--59 & 22 17 57 & --59 41 30 &
93/09/14 & ESO2.2 & EFOSC, 1,8,10 & 10,45,45   \\
59 & RX J2221--27 & 22 21 49 & --27 13 10 &
93/10/14 & ESO2.2 & EFOSC, 1,4 & 10,30   \\
60 & RX J2232--41 & 22 32 43 & --41 34 37 &
93/10/10 & ESO2.2 & EFOSC, 1,8,10  & 10,40,40   \\ 
61 & RX J2241--44 & 22 41 56 & --44 04 55 &
93/10/12 & ESO2.2 & EFOSC, 1,10 & 10,40   \\
62 & RX J2242--38 & 22 42 38 & --38 45 17 &
93/10/11 & ESO2.2 & EFOSC, 1,8,9 & 5,45,45   \\
63 & RX J2245--46 & 22 45 20 & --46 52 12 &
92/10/18 & ESO2.2 & EFOSC, 1,8,9 & 5,30,30   \\
64 & RX J2248--51 & 22 48 41 & --51 09 53 &
92/10/19 & ESO2.2 & EFOSC, 1,8,10 & 5,30,30   \\
65 & MS2254--37 & 22 57 39 & --36 06 07 & 
93/09/12 & ESO2.2 & EFOSC, 1,8,10 & 5,30,30   \\ 
66 & RX J2258--26 & 22 58 45 & --26 09 14 &
92/08/26 & ESO2.2 & EFOSC, 1,8,10 & 15,20,20 
  \\
67 & RX J2301--55 & 23 01 52 & --55 08 31 &
92/10/19 & ESO2.2 & EFOSC, 1,8,10 & ,40,25   \\
68 & RX J2303--55 & 23 03 58 & --55 17 18 &
92/08/24 & ESO2.2 & EFOSC, 1,8,10 & 20,30,30  
  \\ 
69 & RX J2304--35 & 23 04 37 & --35 01 13 &
92/08/25 & ESO2.2 & EFOSC, 1,8,10 & 15,30,30 
  \\
70 & RX J2304--51 & 23 04 39 & --51 27 59 &
93/09/10 & ESO2.2 & EFOSC, 1,8,10 & 5,45,30 
  \\ 
71 & RX J2317--44 & 23 17 50 & --44 22 27 &
93/10/12 & ESO2.2 & EFOSC, 1,8,10 
& 10,40,30   \\
72 & RX J2325--32 & 23 25 12 & --32 36 36 & 
93/10/14 & ESO2.2 & EFOSC, 1,4 & 5,30   \\
73 & RX J2340--53 & 23 40 23 & --53 28 57 &
93/10/14 & ESO2.2 & EFOSC, 1,4 & 5,30   \\
74 & MS2340--15 & 23 43 29 & --14 55 31 &
93/09/14 & ESO2.2 & EFOSC, 1,8,10 & 5,30,30   \\
75 & RX J2349--33 & 23 49 07 & --33 11 45 &
93/10/13 & ESO2.2 & EFOSC, 1,8 & 10,40   \\
76 & RX J2349--31 & 23 49 24 & --31 26 03 &
93/08/21 & ESO2.2 & EFOSC, 1,8,10 
& 30,30,30   \\ \hline
\end{tabular}
\end{flushleft}
\end{table*}

\section{\label{fe2-subspec} FeII subtracted spectra}
In this part of the appendix we present the FeII subtracted spectra. The FeII 
contaminated spectrum and the FeII template are plotted with an offset. The 
upper most spectrum is the original FeII containing one, the middle one shows
the FeII subtracted spectrum and the lower spectrum is the FeII template used 
for the object. A description of the FeII subtraction is given in Sect.
\ref{feii-sub}.

\end{document}